\documentstyle[prl,aps,epsf]{revtex}

\begin{document}
\def \beq {\begin{equation}}
\def \eeq {\end{equation}}
\def \bes {\begin{eqnarray}}
\def \ees {\end{eqnarray}}
\def \ni {\noindent}
\def \nn {\nonumber}
\def \z {\tilde{z}}

\title{
Experimental and Theoretical Investigation of the Lateral 
Casimir Force Between Corrugated Surfaces}

\author{
F.~Chen,
U.~Mohideen\footnote{
E-mail: umar.mohideen@ucr.edu},
}
\address
{Department of Physics, University of California,
Riverside, California 92521\\
}

\author{
G.~L.~Klimchitskaya\footnote{On leave from
North-West Polytechnical University,
St.Petersburg, Russia. E-mail: galina@fisica.ufpb.br}
and V.~M.~Mostepanenko\footnote{On leave from
Research and Innovation Enterprise ``Modus'',
Moscow, Russia. E-mail: mostep@fisica.ufpb.br}
}
\address
{Departamento de F\'{\i}sica, Universidade Federal da Para\'{\i}ba,
C.P.~5008, CEP 58059-970,
Jo\~{a}o Pessoa, Pb-Brazil
}
\maketitle

{\abstract{The lateral Casimir force acting between a sinusoidally
corrugated gold plate and sphere was calculated and
measured.
The experimental setup was based on
the atomic force microscope
specially adapted for the measurement of the lateral Casimir force.
The measured force oscillates sinusoidally as a function
of the phase difference between the two corrugations.
Both systematic and random errors
are analysed and a lateral force amplitude of $3.2\times 10^{-13}\,$N
was measured at a separation distance of 221\,nm with a resulting
relative error 24\% at a 95\% confidence probability. The dependence
of the measured lateral force amplitude on separation was
investigated and shown to be consistent with the inverse fourth
power distance dependence. The complete theory of the lateral
Casimir force is presented including finite 
conductivity and roughness corrections. The obtained
theoretical dependence was analysed as a function of surface
separation, corrugation amplitudes, phase difference, and
plasma wavelength of a metal. The theory was compared
with the experimental data and shown to be in good agreement.
The constraints on hypothetical Yukawa-type interactions
following from the measurements of the lateral Casimir
force are calculated.
The possible applications of the lateral vacuum forces
to nanotechnology are discussed.}}


\section{Introduction}

It is well known that the existence of zero-point electromagnetic
oscillations leads to the Casimir force acting normal
to neutral and parallel metal plates placed in vacuum \cite{1}.
This is a purely quantum effect caused by the alteration of the
zero-point oscillation spectrum of quantized electromagnetic 
field by the metallic boundaries 
(see the monographs \cite{2,3,4} and references therein).
Recently, the normal Casimir force acting perpendicular to the two
surfaces has attracted much experimental and theoretical attention.
It was measured between a flat plate and a spherical lens by means
of a torsion pendulum \cite{5} and between two parallel plates
using a tunneling electromechanical transducer \cite{6}.
The highest precision was achieved in the experimenatal research
of the normal Casimir force between a sphere and a flat plate
by means of the atomic force microscope (AFM) \cite{7,8,9,10}.
In the case of the plate with periodic uniaxial sinusoidal
corrugations and sphere the 
nontrivial boundary dependence of the normal
Casimir force was demonstrated \cite{11}.
These experimental achievements have stimulated an extensive
theoretical study of various corrections to the Casimir force.
Here the finite conductivity
corrections to the normal Casimir force due to the 
boundary metal were investigated in detail
\cite{12,13,14}. The other
influential factor that may contribute considerably to the
normal Casimir force at small separations is surface roughness
\cite{8,15,16,17}. In Refs.~\cite{18,19,20,21,22,23,24} the 
thermal corrections were investigated in the case of real
metals which are significant at separations larger than 
1$\,\mu$m. Also, the combined effect of different corrections
was examined (for a recent review of the subject see
Ref.~{\cite{25}).

There is an important need for further research on the Casimir effect
motivated by the fact that it is finding new applications in
both fundamental science and engineering. Thus, in the framework
of modern unified theories, involving compact extra dimensions
and light elementary particles, precision measurements of the
Casimir force have been used to set limits on the presence
of hypothetical forces \cite{26,27,28,29,30,31}.
Technologically, both static and dynamic micromachines
actuated  by the normal Casimir force have recently been demonstrated
\cite{32,33}. It was also shown that the adhesion and
sticking of moving parts in micromachines 
is due to the Casimir effect \cite{34}.

Similar to the normal Casimir force, the lateral Casimir force may
exist when the bodies are asymmetrically positioned or their
properties are anisotropic. The existence of the lateral
Casimir force opens new opportunities 
for the application of the Casimir force in micromachines.
The lateral Casimir force  also
originates from the modification of electromagnetic zero-point
oscillations by material boundaries. The possibility of a lateral
Casimir force for anisotropic boundaries was investigated
theoretically and a harmonic dependence on a corresponding angle
was predicted \cite{3,35,36}. For two aligned corrugated plates
made of ideal metal the lateral Casimir force was discussed in
\cite{37,38,39} and a harmonic dependence of the result on a
phase shift between corrugations was found. Note that other
motional frictional forces between two flat parallel surfaces
have been suggested but they would be several orders of
magnitude smaller \cite{40,41}.

In Letter \cite{42} the first measurement of the lateral
Casimir force was reported and the theoretical expression
for it was obtained in the case of real metals
of finite conductivity. This
force acts between aligned corrugated 
sphere and a plate in a direction tangential to the 
corrugated surfaces.
The predicted sinusoidal dependence of the lateral force
on the phase shift between corrugations was confirmed.

In this paper we present the detailed experimental and
theoretical investigation of the lateral Casimir force
acting between a corrugated sphere situated near a
corrugated plate with aligned sinusoidal corrugations.
The theoretical dependence for the lateral force
is analysed and the optimum
values of the parameters leading to the maximum values
of the lateral force are found. It is shown that even 
a small misalignment of the corrugation axes will quench 
the lateral Casimir force to zero. The effect of surface
roughness is estimated and shown to be insignificant.
Experimentally, new measurement data are presented
and the calibration procedures by means of the  normal
and lateral electric forces are discussed.
The systematic and random errors 
are analysed and the agreement between theory and experiment 
is confirmed with good precision. The lateral hypothetical force
that may
originate from extra dimensions or from exchange of light 
elementary particles is then calculated and the constraints
on its parameters are obtained.

The paper is organized as follows. In Sec.~II the theory
is developed describing the lateral Casimir force
for the configuration of a metallized sphere and a plate with
the uniaxial corrugations taking into account the finite 
conductivity and roughness
corrections. 
In Sec.~III the experimental setup is described.
Sec.~IV contains the measurement scheme including
calibration procedures.
In Sec.~V the obtained data are presented
together with the error analyses and comparison of the
experimental results to the theory. In Sec.~VI the new
constraints on the parameters of hypothetical interactions
are found.
Sec.~VII contains conclusions and discussion.  

\section{Calculation of the lateral Casimir force between 
surfaces with uniaxial corrugations}

It is well known that the regularized zero-point energy
per unit area for two parallel plates of infinite conductivity
a distance $z$ apart is given by \cite{1,2,3,4,25}
\beq
E_{pp}^{(0)}(z)=-\frac{\pi^2}{720}\,\frac{\hbar c}{z^3}.
\label{1}
\eeq
\ni
This results in the normal Casimir force per unit area
\beq
F_{pp}^{(0)}(z)=
-\frac{\partial{E_{pp}^{(0)}(z)}}{\partial z}=
-\frac{\pi^2}{240}\,\frac{\hbar c}{z^4},
\label{2}
\eeq
\ni
which acts perpendicular to the surface of the plates.

As real metals have only a finite conductivity, corresponding
corrections to Eqs.~(\ref{1}), (\ref{2}) can be incorporated
in terms of the plasma wavelength $\lambda_p$. This was
first done in \cite{43,44,45} up to the first perturbation
order and in \cite{46} up to the second order of a small
parameter $\lambda_p/(2\pi z)$. To get the results applicable
at the separations $z\geq\lambda_p$ with an error of
about (1-2)\%, that are required below, we use the perturbation
expansion up to the fourth order obtained in \cite{12}
(see also \cite{24,25,47}):
\beq
E_{pp}(z)=-\frac{\pi^2 \hbar c}{720{z}^3}
\left[1+
\sum\limits_{n=1}^{4}
c_n\left(\frac{\lambda_p}{2\pi z}\right)^n
\right],
\label{3}
\eeq
\ni
where the coefficients are 
\beq
c_1=-4, \quad c_2=72/5, \quad
c_3=-\frac{320}{7}\left(1-\frac{\pi^2}{210}\right),
\quad
c_4=\frac{400}{3}\left(1-\frac{163\pi^2}{7350}\right).
\label{4}
\eeq

For flat plates at rest, the lateral
Casimir force projection is absent. If, however, the rotational symmetry
against the axis perpendicular to the plates is broken than
the lateral projection of the Casimir force may appear
\cite{3,35,36,37,38,39}. As the first example, let us consider
plates covered by the longitudinal uniaxial corrugations
of equal periods described by the functions
\beq
z_1=A_1\sin (2\pi x/\Lambda),
\qquad 
z_2=z+A_2\sin(2\pi x/\Lambda+\varphi),
\label{5}
\eeq
\ni
where $z$ is the mean separation distance between the two surfaces, 
$\Lambda$ is the 
corrugation period, $A_{1,2}$ are the corrugation amplitudes,
and $\varphi\equiv 2\pi x_0/\Lambda$ is the phase shift
(see Fig.~1).

The normal separation distance between two opposite points of
the corrugated surfaces given by Eqs.~(\ref{5}) is
\beq
z_2-z_1=z+A_2\sin(2\pi x/\Lambda+\varphi)-A_1\sin (2\pi x/\Lambda).
\label{6}
\eeq
\ni
By simple transformations it can be identically represented as
\beq
z_2-z_1=z+b\cos(2\pi x/\Lambda-\alpha),
\label{7}
\eeq
\ni
where the following notations are introduced
\beq
b=b(\varphi)=\left(A_1^2+A_2^2-2A_1A_2\cos\varphi\right)^{1/2},
\qquad
\tan\alpha =(A_2\cos\varphi -A_1)/(A_2\sin\varphi).
\label{8}
\eeq

The representation of the separation distance in the form
of Eqs.~(\ref{7}), (\ref{8}) is convenient for the calculation
of the Casimir energy per unit area between the corrugated plates.
It can be found by additive summation of the results obtained
for plane plates [see Eq.~(\ref{3})]. In doing so we assume  
that all separation distances $z_2-z_1$ given by
Eq.~(\ref{7}) are equally probable. This approximation has been
successfully applied in many calculations of the Casimir
effect in configurations where the variables are not
separable and the exact Green's function of the wave
equation cannot be found explicitly (see, e.g.,
\cite{2,3,8,11,15,16,25}). As was shown in Ref.~\cite{39}
the additive summation works well for corrugated plates
with large corrugation period, e.g. with $\Lambda >z$,
which is the case in our experiment (see Sec.~III).
As a result, the Casimir energy density between
corrugated plates is given by
\beq
E_{pp}^{cor}(z,\varphi)=
\frac{1}{\Lambda}
\int_{0}^{\Lambda}E_{pp}(z_2-z_1)\,dx,
\label{9}
\eeq
\ni
where $z_2-z_1$ is defined by Eqs.~(\ref{7}), (\ref{8}).
Substituting (\ref{7}), (\ref{8}) into (\ref{3}) and
integrating, one obtains
\beq
E_{pp}^{cor}(z,\varphi)=
-\frac{\pi^2\hbar c}{720z^3}
\sum\limits_{n=0}^{4}
c_n\left(\frac{\lambda_p}{2\pi z}\right)^n
X_n(\beta),
\label{10}
\eeq
\ni
where $\beta\equiv b(\varphi)/z$, $c_n$ are defined in 
Eq.~(\ref{4}), and the following notations are used
\bes
&&
X_0(\beta)=\frac{2+\beta^2}{2(1-\beta^2)^{5/2}},\quad
X_1(\beta)=\frac{2+3\beta^2}{2(1-\beta^2)^{7/2}},\quad
X_2(\beta)=\frac{8+24\beta^2+3\beta^4}{8(1-\beta^2)^{9/2}},
\nn \\ 
&&X_3(\beta)=\frac{8+40\beta^2+15\beta^4}{8(1-\beta^2)^{11/2}},
\quad
X_4(\beta)=\frac{16+120\beta^2+90\beta^4+
5\beta^6}{16(1-\beta^2)^{13/2}}.
\label{11}
\ees

Experimentally it is hard to maintain two parallel plates
uniformly separated by distances less than a micron.
So one of the plates is usually replaced by a metallized
sphere or a spherical lens of large radius $R\gg z$
\cite{5,7,8,9,10,11,32,33}. In the experiments described
below (see Secs.III-V) a sphere imprinted with sinusoidal
corrugations was used instead of one of the corrugated
plates. For such a configuration the normal Casimir force
can be calculated approximately by the use of proximity
force theorem (PFT) \cite{48} as
\beq
F^{nor}(z,\varphi)=2\pi R E_{pp}^{cor}(z,\varphi),
\label{12}
\eeq
\ni
where the energy per unit area for the 
configuration of two corrugated
plates is given by the right-hand side of Eq.~(\ref{10}).
For our experimental parameters, 
the two conditions $z\ll R$ and
$\Lambda \ll R$ are fulfiled. As a result the error
introduced by the
PFT in the configuration under consideration is of order
0.2\% \cite{49,50}, which is acceptable for the 
goals of this paper.

By integrating the normal force (\ref{12}) with respect to 
the surface separation,
the energy of a corrugated sphere and a plate is obtained. Then,
differentiating with respect to the phase shift, one finds the
lateral Casimir force
\beq
F^{lat}(z,\varphi)=-\frac{2\pi}{\Lambda}
\frac{\partial }{\partial\varphi}
\int_{z}^{\infty}dy
F^{nor}(y,\varphi).
\label{13}
\eeq
\ni
Substituting Eqs.~(\ref{10})-(\ref{12}) into (\ref{13}) we
finally obtain after integration and differentiation
\beq
F^{lat}(z,\varphi)=\frac{\pi^4 R\hbar c}{120z^4}
\frac{A_1A_2\sin\varphi}{\Lambda(1-\beta^2)^{5/2}}
\left[1+
\sum\limits_{n=1}^{4}c_{n,x}\!\left(
\frac{\lambda_p}{2\pi z}\right)^{\! n}
\right],
\label{14}
\eeq
\ni
where $\beta$ was defined after Eq.~(\ref{10})
 and the expansion coefficients
are given by
\beq
c_{1,x}=\frac{4+\beta^2}{3(1-\beta^2)}c_1,
\quad c_{2,x}=\frac{5(4+3\beta^2)}{12(1-\beta^2)^2}c_2,
\quad
c_{3,x}=\frac{8+12\beta^2+\beta^4}{4(1-\beta^2)^3}c_3,\quad
c_{4,x}=\frac{7(8+20\beta^2+5\beta^4)}{24(1-\beta^2)^4}c_4.
\label{15}
\eeq

The above Eqs.~(\ref{14}), (\ref{15}) give us the expression
for the lateral Casimir force for the 
configuration of a corrugated
sphere and a plate including the finite conductivity
corrections. There are also corrections to Eqs.~(\ref{14}), 
(\ref{15}) due to nonzero temperature. However, at
separations smaller than 0.5$\,\mu$m considered in
Secs.III-V they contribute much less than 1\%
\cite{18,19,20,21,22,23,24,25} and thereby can be neglected.
Another factor that could contribute to the
lateral Casimir force is surface roughness. It was shown to lead
to rather large contributions to the normal Casimir force
at separations below 1$\,\mu$m \cite{15,16,17,25}. Because of
this, the effect of surface roughness on the lateral
Casimir force should be considered in more detail.

There are two kinds of surface roughness on the metal surfaces:
infrequently  distributed tall crystals and short-scale
stochastic distortions. The infrequent tall crystals 
practically do not influence the lateral Casimir force as they are
situated non-periodically and lead to zero contribution
after the averaging over the corrugation period. The situation
here is the same as for two corrugated plates with different
corrugation periods. In Ref.~\cite{15} it was shown that if
the corrugation periods are different (and larger than
a separation distance $z$) the Casimir energy does not depend
on a lateral shift of one plate relative to the other one.
As a result, the derivative of the energy with respect to the 
phase shift is equal to zero and the lateral force is absent.

To take stochastic roughness
into account we can change $F^{lat}(z,\varphi)$
for $F^{lat}(z_i,\varphi)$ with
\beq
z_i=z+x_i,\quad 
\langle x_i\rangle =0,\quad 
\langle x_i^2\rangle = \frac{1}{2}A_{st}^2,
\label{16}
\eeq
\ni
where $x_i$ describes the random change of the separation
distance due to the stochastic roughness with an
amplitude $A_{st}$, and the angle brackets denote the
averaging over the ensemble of all particular realizations
of the corresponding stochastic function. It is important to note
that $z_i$ enters Eq.~(\ref{14}) 
directly as a replacement for $z$ and indirectly
through the
functions of $\beta^2$ that should now be changed to
$\beta_i^2=b^2/(z+x_i)^2$. The lateral Casimir force
with account of stochastic roughness is defined as 
\beq
F_{st}^{lat}(z,\varphi)=
\langle F^{lat}(z_i,\varphi)\rangle.
\label{17}
\eeq
\ni
Performing the computations up to the second order in powers
of $A_{st}/z$ the following result is obtained
\beq
F_{st}^{lat}(z,\varphi)=
\frac{\pi^4R\hbar c}{120z^4}
\frac{A_1A_2\sin\varphi}{\Lambda (1-\beta^2)^{5/2}}
\left[1+\frac{5(1+9\beta^2-3\beta^4)}{4(1-\beta^2)^2}
\frac{A_{st}^2}{z^2}\right]\,
\left[1+
\sum\limits_{n=1}^{4}
c_{n,x}^{st}\left(\frac{\lambda_p}{2\pi z}\right)^n\right].
\label{18}
\eeq
\ni
Here the coefficients $c_{n,x}^{st}$ are only slightly
different from those given by Eq.~(\ref{15}) (which does not 
include stochastic roughness). For example,
for $n=1,\,2$ their expressions are
\beq
c_{1,x}^{st}=c_{1,x}\left[1+
\frac{15\beta^2}{2(4+\beta^2)(1-\beta^2)}
\frac{A_{st}^2}{z^2}\right],
\qquad
c_{2,x}^{st}=c_{2,x}\left[1+
\frac{3\beta^2(11+3\beta^2)}{2(4+\beta^2)(1-\beta^2)}
\frac{A_{st}^2}{z^2}\right].
\label{19}
\eeq
\ni
If we take into consideration the typical values of
$\beta^2<0.1$ and $A_{st}\approx 10\,$nm, Eq.~(\ref{18})
can be approximately rewritten in a more simple form
\beq
F_{st}^{lat}(z,\varphi)\approx
F^{lat}(z,\varphi)
\left[1+\frac{5(1+9\beta^2-3\beta^4)}{4(1-\beta^2)^2}
\frac{A_{st}^2}{z^2}\right].
\label{20}
\eeq
\ni
{}From Eq.~(\ref{20}) one can conclude that at separations
$z>200\,$nm used in the experiment the influence of
stochastic roughness on the lateral force is less than
1\% and can be neglected. Thus, Eqs.~(\ref{14}), (\ref{15})
give us a reliable theoretical expression for the
lateral Casimir force including all necessary corrections.

The most interesting characteristic feature of Eq.~(\ref{14})
is the harmonic dependence of the lateral Casimir force on
a phase shift between the corrugations of both bodies.
However, the actual dependence of $F^{lat}$ on $\varphi$
is not exactly sinusoidal because $\beta$ also depends
on $\varphi$ which leads to some deviation from the exact
sine. To illustrate this, in Fig.~2 the dependence
of  $F^{lat}/F_{max}^{lat}$ on $\varphi$ at a separation
$z=272\,$nm is plotted (solid line). In the same figure
the graph of $\sin\varphi$ is shown by a dashed line.
To make deviations from a perfect sine larger, the case
of equal amplitudes $A_1=A_2=59\,$nm is considered
(in the experiment of Sec.III-V the amplitude of corrugations
on a sphere is smaller than on the plate).
As is seen from Fig.~2, the maximum of the lateral Casimir
force is displaced from the position of the maximum of sine
by approximately 0.21\,rad. 

The values of the lateral force given by Eq.~(\ref{14}) depend
on the corrugation amplitudes (both in an explicit form and
through the parameter $b$). In Fig.~3 the graph of $F_{max}^{lat}$ 
as a function of $A_2$ is plotted for 
$A_1=59\,$nm. For each
value of $A_2$ the distance $z=z_0+A_1+A_2$ is chosen where
$z_0=154\,$nm that is in accordance with the experimental
value of the separation on contact (see Secs.III-V).
It is seen that $F_{max}^{lat}$ increases with an increase of $A_2$
and takes the largest value $F_{max}^{lat}=1.2\times 10^{-12}\,$N
when $A_2=A_1$.

The effects of the finite conductivity of the boundary metal 
makes a significant 
contribution to the value of the lateral Casimir force
from Eq.~(\ref{14}). This is illustrated by Fig.~4, where the 
correction coefficient $\eta=F^{lat}/F_0^{lat}$ is plotted as
a function of separation distance, and $F_0^{lat}$ is computed
for an ideal metal (i.e. with $\lambda_p=0$).
Here the experimental values of the corrugation amplitudes were
chosen, i.e. $A_1=59\,$nm, $A_2=8\,$nm, and a phase shift
$\varphi=\pi/2$ (see Secs.III-V). It should be noted that the value 
of the correction factor $\eta$ depends only slightly on the
phase shift. The value of the plasma wavelength
$\lambda_p=136\,$nm for $Au$ was used \cite{13}. 
It is seen from the figure that in the separation
range of interest here the correction coefficient changes between 
0.6 and 0.7. Because of this, it would be incorrect to use
a theory which does not include effect  
of the finite conductivity corrections for
interpretation of the experimental data on the lateral Casimir
force.

At the end of this section we briefly discuss the demand  
that the corrugations be uniaxial. This demand is of crucial
importance for the observation of the lateral Casimir force. In fact,
let us assume for a moment that there is some nonzero angle
$\vartheta$ between the corrugation axes. Then the phase
shift $\varphi$ along the $x$-axis becomes the periodical
function of $y$ with a period 
$\Lambda_y=\Lambda\cot\vartheta$. In the limit of one
period $\varphi(y)$ depends on $y$ linearly, taking on values from
0 to $2\pi$. To obtain the resulting lateral force, the
expression $F^{lat}\left(z,\varphi(y)\right)$ should be averaged 
over the period $\Lambda_y$ which leads to a zero value.
For real bodies of finite size the
lateral Casimir force will exist  
only for small deviations of the corrugation
axes from parallelity such 
that $\Lambda\cot\vartheta$ is much
larger than the smallest body. In our case the smallest body
is the 10-micron section of the sphere covered with corrugations. 
That is why in order 
to observe the lateral Casimir
force one must make sure that the angle between the corrugation
axes is bounded by the condition $\vartheta\ll 0.1\,$rad.

\section{Experimental setup}

A schematic diagram of
the experiment is shown in Fig.~5. These experiments 
are performed using a standard AFM
at a pressure below 50\,mTorr and at room temperature. 
The experiment requires 
two sinusoidally corrugated surfaces with their respective 
axes perfectly parallel. 
Misallignment by 3${}^{\circ}$ of the 
corrugation axes can lead to loss of any lateral force
due to the cross-over of the 
axes as is described in Sec.II.   A plastic diffraction 
grating with an uniaxial sinusoidal corrugations of period
$\Lambda=1.2\,\mu$m
and an amplitude of 90\,nm was used as the 
corrugated plate. 
In order to
obtain perfect orientation and phase between the two corrugated 
surfaces a special {\it in situ} procedure was developed, where  
the corrugations from the plate are imprinted on the gold coated 
sphere by pressure. This imprinting procedure required special 
adaptation of the
cantilever which is described next.  

A polystyrene sphere was attached to the tip 
of a 320 \,$\mu$m long cantilever with conductive silver epoxy. After 
this a $<10\,\mu$m thick, $100-200\,\mu$m wide and 0.5\,mm long piece
of freshly
cleaved mica is attached to the bottom of the sphere 
with silver
epoxy. Then a second polystyrene sphere of $2R=200\pm4\,\mu$m
diameter was mounted on the tip of
mica with the same silver epoxy.  This second sphere is 
imprinted with the corrugations and will interact with 
the corrugated plate.   The sphere and the
plates are mounted as shown in Fig.~5.  
Let us first note that the laser beam for the 
detection of the cantilever deflection is reflected off its tip. 
The addition of the first sphere and mica plate is needed to isolate 
the laser reflection spot on the cantilever tip 
from the interaction region between the two 
corrugated surfaces.  This isolation is necessary to reduce 
the effect of scattered light from the top and sides   
of the corrugated plate.  Secondly, the 
procedure developed for the imprinting of the corrugations 
requires access to interior regions of the 
corrugated plate, far away from 
the edges.   Thirdly, the addition of the mica plate 
leads also 
to an effective increase in the detection sensitivity 
due to the increase in the lever arm. 
The cantilever (with mica plate and 
spheres), corrugated plate  and a smooth flat plate (polished sapphire) 
were then coated with about 400 nm of gold in a thermal
evaporator.  A small region close to one edge of the 
corrugated plate is also coated with
100\,nm of aluminum. As $Al$ exhibits more hardness 
than gold, this region is used to imprint the corrugations from 
the plate onto the gold coated sphere.

The cantilever (with mica plate and 
spheres), corrugated plate  and a smooth 
flat plate are 
then mounted as shown in Fig.~5. 
Now, the imprinting of the 
corrugations on the sphere is done. The sphere is 
moved over to the region of the
corrugated plate coated with $Al$.  The other side of the 
sphere is mechanically supported and the corrugations
are imprinted on the gold
coating of the surface by pressure using the piezos shown.  A 
scanning electron micrograph of the imprint on the sphere, taken 
after the completion of the experiment, is 
shown in Fig.~6. An AFM scan of the imprinted corrugations is 
shown in Fig.~7.  The amplitude of the imprinted corrugations is 
measured from the AFM scan to be  $A_2=8\pm1\,$nm .  The amplitude of
the corrugations on the metallized plate was
also measured, using the AFM, 
$A_1=59\pm 7\,$nm. These AFM measurements were made 
after completion of all the lateral force experiments 
which are reported below. 
After this imprinting, the mechanical supports are removed and 
the sphere is
translated  over to the gold coated area of the plate. Extreme 
care to preserve the parallel orientation of the two corrugations 
is necessary during this translation, as any misallignment leads to 
the destruction of any lateral Casimir force. This is done 
by tracking the orientation of the cantilever during this 
translation by reflecting two
optical beams from the edges of the 
cantilever holder.  The reflected 
beam positions  allow measurement of the 
cantilever orientation to an 
accuracy of 2$\times 10^{-3}$\,rad. 

The corrugated 
plate is mounted on two piezo electric tubes 
that allow independent movement of the 
plate in the vertical and the horizontal directions 
with the help of a $x$-piezo and 
a $z$-piezo respectively. Movement in the 
$x$ direction with the $x$-piezo is necessary to achieve lateral phase 
shift $\varphi$ 
between the 
corrugated sphere and the plate.  Independent movement in the  
$z$ direction is necessary for control of the surface separation between 
the corrugated sphere and plate. The corrugated plate is 
mounted vertically 
in order to increase the 
sensitivity for detection of lateral 
forces and suppress the effect of the 
normal Casimir force on the cantilever. Thus a 
lateral force tangential 
to the corrugated sphere surface 
would
result in the usual bending of the cantilever in response 
to the force. Whereas a force acting normal to the 
sphere and corrugated plate (from the normal Casimir force) 
would lead to the torsional deflection (rotation) 
of the cantilever. 
The torsional spring constant of this cantilever $k_{tor}$ 
is much greater than the bending spring constant $k_{ben}$, 
making it much more sensitive to detecting the 
lateral Casimir force, while simultaneously 
suppressing the effect of the normal Casimir force. 

\section{Measurement scheme}

The calibration of the cantilever 
($k_{tor}$ and $k_{ben}$)
and the measurement of the residual potentials between the sphere and plate 
is done by electrostatic means ~\cite{7,8,9,10,11}. These calibrations are 
done after the measurement of the lateral Casimir forces, but is reported
in this section for the benefit of continuity. Here, in order to measure 
$k_{tor}$, the sphere is kept grounded and various 
voltages are applied to the corrugated plate.  
The normal electrostatic force
between the corrugated sphere and plate is given by:
\beq
F_{z}^{el}(z,\varphi)=-\pi R\varepsilon_0
\frac{(V_1-V_0)^2}{z}
\frac{1}{\sqrt{1-\beta^2}},
\label{21}
\eeq
\ni
where $\varepsilon_0$ 
is the permittivity of free space. $V_1$ are the voltages 
applied on the corrugated plate and $V_0$ 
is the residual potential on the grounded sphere. 
The approximate expression (\ref{21}) was obtained by exactly
the same procedure as Eq.~(\ref{12}) for the normal Casimir
force
[instead of Eq.~(\ref{3}),
we have started here from the energy per unit area of a capacitor
formed by two large, flat conducting sheets].

If $V_1$ is applied 
to the corrugated plate, the  
electrostatic force acting normal to the 
spherical surface  
leads to the torsional rotation of the cantilever. By applying 
different $V_1$ we can solve for the torsional spring constant 
$k_{tor}=0.138\pm 0.005\,$N/m and the residual voltage between the 
sphere and the corrugated plate ${V_0}=-0.135\,$V.  Next the 
measurement of $k_{ben}$ is done.  The sphere is 
moved away from the vertical corrugated plate and brought closer 
to the smooth plate which is positioned 
horizontally at the bottom as shown in Fig.~5. 
Again different voltages $V_1$ are
applied to the bottom plate 
[here in Eq.~(\ref{21}) $ A_{1}=A_{2}=\beta=0$ 
due to the smooth surfaces], 
and the electrostatic force 
leads to the normal bending of the cantilever. 
We again solve for the normal spring constant 
$k_{ben} =0.0052\pm 0.0001\,$N/m and the residual voltage 
between the sphere and the smooth plate.  
Note that $k_{tor}\gg k_{ben}$ is required 
for isolation and detection of the role of the lateral Casimir force. 
The piezo extension in the $x$ direction
with applied voltage was calibrated by optical interferometry \cite{50a}.
The horizontal displacement of the piezo in the $z$ direction was 
calibrated with AFM standards.

Similar to the lateral Casimir force, there also exists a lateral 
electrostatic force which arises from the presence of an 
applied or residual electrostatic potential difference 
between the two corrugated surfaces. It is given by
\beq
F_{x}^{el}(z,\varphi)=2\pi^2 R\varepsilon_0
{(V_1-V_0)^2}\frac{A_1A_2}{\Lambda z^2}
\frac{\sin\varphi}{\sqrt{1-\beta^2}(1+\sqrt{1-\beta^2})}.
\label{22}
\eeq
\ni
This expression is obtained from Eq.~(\ref{21}) by integration
with respect to $z$ (in order to find the electric energy in
configuration of a corrugated plate and a corrugated sphere)
and differentiation with respect to $x_0=\varphi\Lambda/2\pi$. 
Both Eqs.~(\ref{21}),
(\ref{22}) are valid with an error smaller than 1\% for the
experimental parameters under consideration.

In contrast with the lateral Casimir force, the  
lateral electrostatic force 
is dependent on the inverse second power of the separation 
distance $z$ between the corrugated surfaces (the dependence
of $\beta$ on $z$ is small). One can 
measure the lateral electrostatic force in order to distinguish 
its differences from the lateral Casimir force. The measurement 
of the lateral electrostatic force also will help in 
providing an approximate measure of the 
separation distance between the two corrugated surfaces on contact. 
Note that because of the roughness of the metal 
surfaces and the imprinting procedure used, 
the contact separation is much greater than the 
distances between the means of the corrugations.

Two different voltage differences 
between the corrugated plate and 
sphere were used in the measurements of the lateral 
electrostatic force. In the first case, we 
utilized the residual voltage 
difference $V_0=0.135\,$V with $V_1=0\,$V.  The sphere was moved 
next to the corrugated plate and the separation distance 
between the two surfaces was kept fixed.
To measure the 
lateral electrostatic  force $ F_x^{el}$ as a function 
of the phase $\varphi$, the corrugated 
plate is moved in the $x$-direction by $x$-piezo
in average steps of  0.46\,nm  
(due to the small nonlinear response of the piezo, 
the exact step size will differ by a few percent depending 
on the applied voltage \cite{50a}) 
and the lateral electrostatic force is measured at each step.
The corrugated plate could have been mounted with a small
but nonzero tilt away from the vertical ($x$-axis). Such tilts would
lead to changes in surface separations during the above
translations of the plate in the $x$ direction. In order to
rectify this, a small correction voltage is applied to the
$z$-piezo, synchronous with the lateral translation in
$x$ direction, to keep the surface separation distance between
the corrugated sphere and plate constant. The lateral force 
measurement is repeated 60 times and the average 
lateral force at each step is recorded. 
The measured force in this case is  
actually the sum of the lateral 
Casimir force and the lateral electrostatic force. 
The observed lateral force is shown in Fig.~8.
A sine curve is best fit to the observed data 
and an amplitude of 
$16.2 \times 10^{-13}$N is obtained for the 
total force. This amplitude when fit to 
the sum of the two lateral
forces (Casimir+electrostatic) 
resulting from Eq.~(\ref{14}) and 
from Eq.~(\ref{22}) leads to a separation distance of 
$z=225\pm 4\,$nm between the two corrugated surfaces. 
This separation distance is used to subtract the lateral Casimir 
force from the measured total force to obtain the 
lateral electrostatic force. The error in the separation 
distance $z$ corresponds to the error in the amplitude of the 
electrostatic force resulting from the 16nm uncertainty 
in $x$. This uncertainty is $x$ was determined 
experimentally by measuring the random variations in 
the phase of the 
peaks of the sinusoidal oscillations from 
30 scans. Note that this random uncertainty in the 
phase corresponding to 16nm is much smaller than 
the period of corrugations ($\Lambda=1.2\,\mu$m).  
Next, the separation between the sphere and 
corrugated plate is decreased by 24\,nm 
and the measurement is repeated. This was repeated till 
the sphere and the corrugated plate come in 
contact.  The 
surface separation on contact of the 
two corrugated surfaces is $202\pm$38\,nm (the 
large uncertainty is total of the 
uncertainty of 24\,nm resulting from the 
step size, the 5\,nm systematic uncertainty from the 
measurement of the force amplitude and the 9\,nm random 
error from the force measurement at different separation
distances).

In the second case
 a different voltage $V_1= -0.055\,$V is applied to the 
corrugated plate ($V_0=-0.135\,$V) and the lateral electrostatic 
force measurement is repeated. Again the distance 
between the corrugated surfaces is changed in steps of 
24\,nm, starting at some separation, till the 
two surfaces come into contact. The separation between 
the two corrugated surfaces on contact was 
$169\pm$33\,nm (in this case a 4\,nm random error from the force
measurement at different separations was present).  
Thus the average separation on contact 
from the two applied voltages was $186\pm$38\,nm. 
Note that the lateral electrostatic force measurement was 
done a few hours after the measurement of the lateral 
Casimir force (described below) and thus this separation 
distance on contact obtained from the lateral 
electrostatic force serves as only a constraint on the 
separation distances between the corrugated surfaces 
to be expected for the measurement of the lateral 
Casimir force. 
In Fig.~9, a log${}_{10}$-log${}_{10}$ plot of the measured 
lateral electrostatic force amplitude as a 
function of the separation distance is done 
for the two
applied voltages as solid squares and 
triangles respectively.   
The slopes of the best straight fit lines (using the 
least squares procedure) to the 
two sets of measured lateral electrostatic forces are 
2.5$\pm$0.4 and 2.0$\pm$0.3, respectively, leading 
to an average slope of 2.3$\pm$0.4. Thus 
the measured slope is consistent with the 
second power distance dependence expected from 
Eq.~(\ref{22}).

\section{Obtained data, error analyses, and comparison
with theory}

In this section we discuss the results obtained with the 
measurement of the lateral Casimir force. The measurement procedures 
described above with the measurement of the lateral 
electrostatic force are used. The important difference is that 
for the measurement of 
the lateral Casimir force, the residual potential 
difference between the corrugated sphere and plate is compensated 
by application of voltage $V_0$ to the corrugated plate.  
As before the sphere is brought close to the corrugated plate and the 
separation distance is kept fixed. To measure the 
lateral Casimir force $ F^{lat}$ as a function 
of the phase $\varphi$, for a given sphere-plate separation, 
the corrugated 
plate is moved in the $x$-direction in average steps of 0.46\,nm  
using the 
$x$-piezo and the lateral Casimir force is measured at each step.
As discussed above, correction voltages are applied to the
$z$-piezo synchronous with the movement in the $x$-direction
to correct for any tilts from the mounting of the corrugated plate
away from the vertical ($x$-axis). 
This is repeated 60 times and the average 
lateral Casimir force at each step is recorded. The average 
lateral Casimir force measured is shown as the 
solid squares in Fig.~10.  The scattered laser light leads 
to a small linear drift. Thus a corresponding straight line 
has been subtracted from the acquired data.  The
sinusoidal oscillations in the lateral Casimir force 
expected from Eq.~(\ref{14}) as a 
function of the phase difference between the two 
corrugations are clearly observed. The periodicity 
of the lateral Casimir force oscillation is also in agreement with 
corrugation period of the plate.   
A sine curve fit to the observed data is shown as the 
solid line and corresponds to an amplitude of 
3.2$\times 10^{-13}$N. 
{}From Eq.~(\ref{14}) this corresponds to a separation distance of 
$z=221\pm$2\,nm between the two corrugated surfaces.

Here a more detailed error analyses is performed in 
comparison to Ref.\cite{42}.
The mean quadratic error of the average lateral force amplitude
is $\sigma_{\bar{A}}=0.22\times 10^{-13}\,$N. The largest 
source of the systematic error is due to the resolution of the 
A/D board used in the data acquisition.  This systematic error
is $\Delta_{A}^{(s)}=0.33\times 10^{-13}\,$N. Using the value of Student
coefficient $t_{0.95,\,60}=2$ one obtains for the 
half-width of the confidence interval, or for the total absolute error, 
$\Delta_{A}=\Delta_{A}^{(s)}+2\sigma_{\bar{A}}=0.77\times 10^{-13}\,$N 
with a 95\% confidence
probability. The resulting precision of the amplitude measurement 
at the closest point is around 24\%.
 
The above lateral Casimir force measurement is repeated for other surface 
separations. First, the separation between the sphere and 
corrugated plate is increased by 12\,nm with the $z$-piezo 
and the measurement is repeated. 
The average measured amplitude of 
lateral force is $2.6\times 10^{-13}\,$N. Based on  
Eq.~(\ref{14}) this corresponds to $z=233\pm$2\,nm 
consistent with the 12\,nm increase in 
the separation distance. Thus the measured lateral Casimir force 
is in agreement with the complete theory taking into account 
the conductivity corrections. 
The separation distance is  
increased in 12\,nm steps and the lateral Casimir force is 
measured for two more surface separations. The amplitudes of the measured 
forces $2.1\times 10^{-13}\,$N and $1.7\times 10^{-13}\,$N 
were found to be consistent with the corresponding separation 
distances.  In Fig.~11 a log${}_{10}$-log${}_{10}$ plot of the 
amplitudes of the measured lateral force 
as a function of the various separation distances is shown as 
solid squares.  Here the separation distance of 221\,nm 
determined from Fig.~10 is used for the closest point. 
The remainder of the points are fixed by the 12\,nm step 
increase in the separation distance.  
A linear fit to the data yields a 
slope of 4.1$\pm$0.2 consistent with the inverse
fourth power $z$ dependence of the lateral force 
expected from Eq.~(\ref{14}).  Note that the 
corrections to this fourth power 
dependence are rather small given that the  
value $\beta<0.3$.
Thus, the lateral Casimir force demonstrates a very  
different dependence 
on separation distance than the 
lateral electrostatic force which leads to an 
inverse second power  $z$ dependence.

\section{Lateral hypothetical forces and constraints
on their parameters}

The measurement of the lateral Casimir force presented in the
above section gives the possibility to constrain the parameters
of the hypothetical long-range interactions which may act 
between the test bodies.
The problem of hypothetical long-range interactions has a long 
history. It is well known that such interactions complementary
to the gravitational and electromagnetic forces are predicted
by many extensions to the Standard Model \cite{51}. They may
be caused by the exchange of light elementary particles
\cite{3,25} or by
extra-dimensional physics
with a low compactification scale \cite{52}. In both cases, 
additional Yukawa-type interactions are predicted that can be
described by the potential
\beq
V^{Yu}(r)=
-\frac{Gm_1m_2}{r}\left(1+\alpha_Ge^{-r/\lambda}\right),
\label{6.1}
\eeq
\ni
where $G$ is the Newtonian gravitational constant, $m_{1,2}$
are the masses of the atoms, $r$ is the separation distance
between them, $\alpha_G$ is the dimensionless constant of
hypothetical interaction, and $\lambda$ is the interaction range.

It is common knowledge that at $\lambda >10^{-4}\,$m
the gravitational experiments of E\"{o}tvos- and Cavendish-type
lead to the strongest constraints on $\alpha_G$ \cite{51}.
However, for smaller $\lambda$ the best constraints on
$\alpha_G$ were obtained from the measurements of the normal
Casimir force \cite{26,27,28,29,30,31}. The above measurements
of the lateral Casimir force deal with smaller forces than
the previous experiments on the normal force. Thus
they may lead to the competitive
constraints on hypothetical interaction.

We start with the calculation of the lateral hypothetical
Yukawa-type interaction for the configuration of a plate and a sphere
with uniaxial corrugations. The same procedure as was used
above for the Casimir force is applicable.
The hypothetical interaction between two flat parallel plates
of mass densities $\rho$ and $\rho^{\prime}$ covered by a thin
$Au$ layer of density $\rho_{Au}$ and thickness $\Delta$
can be obtained by an additive summation
of the Yukawa parts of interatomic potentials (\ref{6.1}). 
The result is \cite{3,25,26,27,29,31}

\beq
E_{pp}^{Yu}(z)=-2\pi G\alpha_G\lambda^3e^{-z/\lambda}
\times\left[\rho_{Au}-\left(\rho_{Au}-\rho\right)
e^{-\Delta/\lambda}\right]\,
\left[\rho_{Au}-\left(\rho_{Au}-\rho^{\prime}\right)
e^{-\Delta/\lambda}\right].
\label{6.2}
\eeq
\ni
Note that in Sec.II, where the Casimir force was calculated, only
the top metallic $Au$ covering layers were essential and the underlying
substances did not influence the force value.

The corrugations on both plates can be included by changing $z$ in
Eq.~(\ref{6.2}) for $z_2-z_1$ defined in Eq.~(\ref{7}) and by
averaging the obtained quantity over the corrugation period
in accordance with Eq.~(\ref{9}). The result is given by
\beq
E_{pp}^{cor,Yu}(z,\varphi)=E_{pp}^{Yu}(z)I(\varphi),
\label{6.3} 
\eeq
\ni
where the notation is introduced
\beq
I(\varphi)\equiv\frac{1}{2\pi}
\int_{0}^{2\pi}dt e^{-\frac{b(\varphi)}{\lambda}\cos(t-\alpha)},
\label{6.4}
\eeq
\ni
$b(\varphi)$ and $\alpha$ are defined in Eq.~(\ref{8}).

Using PFT from Eq.~(\ref{12}) and integrating the obtained
force with respect to the separation distance, one finds
the energy in the configuration of a corrugated plate and a sphere
as
\beq
E^{Yu}(z,\varphi)=2\pi R\lambda 
E_{pp}^{Yu}(z)I(\varphi)=2\pi R\lambda
E_{pp}^{cor,Yu}(z,\varphi).
\label{6.5}
\eeq
\ni
Differentiating Eq.~(\ref{6.5}) with respect to a phase shift 
as it was done in Eq.~(\ref{13}) we come to the expression for
the lateral hypothetical force for the configuration of a plate
and a sphere covered with uniaxial corrugations
\beq
F^{lat,Yu}(z,\varphi)=-\frac{4\pi^2R\lambda}{\Lambda}
E_{pp}^{Yu}(z)\frac{dI(\varphi)}{d\varphi}.
\label{6.6}
\eeq
\ni
The derivative with respect to $\varphi$ can be calculated most
easily if one uses the representation of the quantity $I$ from
Eq.~(\ref{6.4}) in the form of an infinite series
\beq
I(\varphi)=1+
\sum\limits_{n=1}^{\infty}\frac{a_n}{(2n)!}
\left[\frac{b(\varphi)}{\lambda}\right]^{2n},
\label{6.7}
\eeq
\ni
where
\beq
a_n\equiv\frac{1}{2\pi}
\int_{0}^{2\pi}dt (\cos t)^{2n}.
\label{6.8}
\eeq
\ni
Differentiating Eq.~(\ref{6.7}) with respect to $\varphi$
along with use of Eq.~(\ref{6.8}) and substituting into
Eq.~(\ref{6.6}) one finally obtains the lateral
hypothetical force in the form
\beq
F^{lat,Yu}(z,\varphi)=-4\pi^2RE_{pp}^{Yu}(z)
\frac{A_1A_2}{\Lambda b(\varphi)}\sin\varphi
\sum\limits_{n=1}^{\infty}\frac{a_n}{(2n-1)!}
\left[\frac{b(\varphi)}{\lambda}\right]^{2n-1},
\label{6.9}
\eeq
\ni
where $E_{pp}^{Yu}(z)$ is defined in Eq.~(\ref{6.2}).
Notice that the coefficients $a_n$ from Eq.~(\ref{6.8})
are simply calculated (e.g., $a_1=0.5$, $a_2=0.375$,
$a_3=0.3125$ etc) and the sum converges rapidly
due to the factorial terms.

Now we are in a position to find the constraints on the
hypothetical interactions following from the measurements
of the lateral Casimir force. They can be obtained from
the inequality
\beq
\left\vert F_{max}^{lat,Yu}\right\vert <\Delta_A,
\label{6.10}
\eeq
\ni
where $\Delta_A=0.77\times 10^{-13}\,$N is the total absolute error
of the lateral Casimir force measurements with a 95\%
confidence probability (see Sec.V). The quantity 
$F_{max}^{lat,Yu}$ is the maximal value of the lateral
hypothetical force from Eq.~(\ref{6.9}) with respect to a
phase shift $\varphi$ computed at a smallest separation
distance $z=221\,$nm (note that the lateral Newtonian force is many
orders less than $\Delta_A$). 
The obtained constraints are plotted in
Fig.~12 by the solid curve. The region of
($\lambda,\alpha_G$)-plane above the curve is prohibited, and
below the curve is permitted by the results of the measurement of
the lateral Casimir force. 
$\lambda$ is measured in meters and the logarithm to the base 10
is used.
The short-dashed curve in Fig.~12
was obtained from the old Casimir force measurements between
dielectrics (see \cite{3,25}). The long-dashed curve follows
\cite{31} from the measurements of the normal Casimir force
between gold surfaces by means of atomic force microscope
\cite{10}. It is seen that in the interaction range
$80\,\mbox{nm}<\lambda<150\,$nm the constraints obtained by means
of the lateral Casimir force measurements are of almost the
same strength as the previous results. They, however, 
can be considered as more
reliable as in the Casimir force measurements between dielectrics 
the measurement error was estimated rather approximately, whereas in
Refs.~\cite{10,31} the confidence level and confidence probability
were not indicated. 

\section{Conclusions and discussion}

In the above, the experimental and theoretical investigation of the
lateral Casimir force is presented. 
The lateral Casimir force was
first demonstrated in Ref.~\cite{42}. Here 
the measurements were performed with the use of
an AFM specially adapted to increase the sensitivity
for detection of the lateral Casimir forces. The measured
lateral force has the periodic dependence on the phase
shift between the corrugations on both test bodies. The period
of the lateral force coincides with the period of
corrugations. The amplitude of the lateral force was found
to be equal to $3.2\times 10^{-13}\,$N at the separation
distance 221\,nm. 
The resulting experimental
relative error of the amplitude measurement is 24\% with
a 95\% confidence probability.

The normal electrostatic force between a sphere and a plate 
was used for both calibration of the cantilever and for
the measurement of the residual potentials between the test 
bodies. The lateral electrostatic force leading to the
inverse second power distance dependence is applied
for the independent measurement of surface separation
for the first time.
The inverse fourth power dependence of the lateral Casimir
force on separation distance was confirmed with high
precision.

The experimental data were compared with
a complete theory taking into account both finite conductivity
and roughness corrections to the lateral Casimir force 
(the temperature corrections are not important at separations
smaller than 0.5\,$\mu$m). The finite
conductivity corrections to the lateral Casimir force 
decrease the result computed for ideal
metals by 30-40\% in the separation range under
consideration. Thus, the inclusion of these
corrections is necessary for the comparison of theory and
experiment.  

The obtained experimental data on the lateral Casimir force
were used to set constraints on the constants of
Yukawa-type hypothetical interactions.
In the interaction range 
80\,nm$<\lambda <150\,$nm the obtained constraints are
shown to be quite competitive (although a bit weaker)
with the previously known ones from the measurement
of the normal Casimir force. 
In future with the increased precision one
may expect that stronger constraints on the parameters
of hypothetical long-range interactions will be obtained
from the measurements of the lateral Casimir force.

Another prospective application where the above results
can be used is in nanotechnology. With device dimensions 
shrinking to hundreds and even to tens nanometers 
the Casimir force becomes the leading force which
determines its functioning. 
The existence of the lateral Casimir force
in the case of corrugated surfaces gives the possibility
to actuate both normal and lateral translations by means
of the electromagnetic zero-point fluctuations. This
opens new promising opportunities for the application
of the Casimir effect in microelectromechanical systems.

\section*{Acknowledgments}
 
This work is supported by a National Science Foundation Nanoscale 
Exploratory Research Grant and
the National Institute 
for Standards and Technology through a Precision 
Measurement Grant. G.L.K. and V.M.M. were also supported 
by CNPq.


\newpage
\noindent
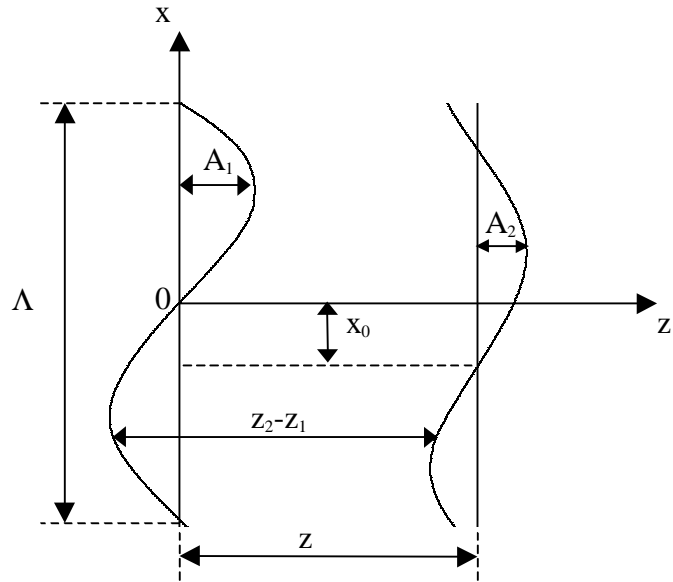
{\bf Figure 1:} Configuration of two parallel plates with
uniaxial sinusoidal corrugations of equal periods.
\hfill\\[5mm]
{\bf Figure 2:} The lateral Casimir force between the corrugated
plate and sphere normalized for its maximum value as a function
of a phase shift (solid line) is compared to a graph of
sine (dashed line).
\hfill\\[5mm]
{\bf Figure 3:} The maximum value of the lateral Casimir force 
as a function of a corrugation amplitude on a sphere.
\hfill\\[5mm]
{\bf Figure 4:} Correction coefficient due to the effects of
finite conductivity on the lateral Casimir
force between the corrugated plate and sphere made of ideal
metals as a function of surface separation.
\hfill\\[5mm]
{\bf Figure 5:}  Schematic of experimental 
setup. For clarity, the sizes of the 
corrugations have been exaggerated. 
The $x$-piezo and $z$-piezo are independent. 
\hfill\\[5mm]
{\bf Figure 6:} Scanning electron micrograph of the imprint
of the corrugations on the sphere.
\hfill\\[5mm]
{\bf Figure 7:} Atomic force microscope scan of the imprinted
corrugations on the sphere.
\hfill\\[5mm]
{\bf Figure 8:} The average measured sum of the electric and
Casimir lateral forces as a function of the lateral 
displacement of the corrugated plate is 
shown as solid squares. The solid line 
is the best fit sine curve to the data
leading to a lateral force amplitude 
of $16.2\times 10^{-13}\,$N.
\hfill\\[5mm]
{\bf Figure 9:} The log${}_{10}$-log${}_{10}$ plot of the measured 
lateral electrostatic force amplitude as a 
function of the surface separation 
distance  is shown as solid squares. 
\hfill\\[5mm]
{\bf Figure 10:}  The average measured lateral 
Casimir force as a function of the lateral 
displacement of the corrugated plate is 
shown as solid squares. The solid line 
is the best fit sine curve to the data 
leading to a lateral force amplitude 
of $3.2\times 10^{-13}\,$N.
\hfill\\[5mm]
{\bf Figure 11:}  The log${}_{10}$-log${}_{10}$ plot of the measured 
lateral Casimir force amplitude as a 
function of the surface separation 
distance  is shown as solid squares. 
The slope of the straight line fit is 4.1$\pm$0.2.
\hfill\\[5mm]  
{\bf Figure 12:} Constraints on the Yukawa-type hypothetical 
interactions following from the measurement of the lateral
Casimir force between corrugated surfaces (solid curve),
normal Casimir force between gold plate and a sphere
(long-dashed curve), and normal Casimir force between
dielectrics (short-dashed curve). The logarithm is to the
base 10.

\begin{thebibliography}{99}
\bibitem {1}
H.~B.~G.~Casimir,
{ Proc. K. Ned. Akad. Wet.}
{\bf 51}, 793 (1948).
\bibitem{2}
P.~W.~Milonni,
{\it The Quantum Vacuum}
(Academic Press, San Diego, 1994).
\bibitem{3}
V.~M.~Mostepanenko and N.~N.~Trunov,
{\it The Casimir Effect and its Applications}
(Clarendon Press, Oxford, 1997).
\bibitem{4}
K.~A.~Milton, {\it The Casimir Effect}
(World Scientific, Singapore, 2001).
\bibitem{5}
S.~K.~Lamoreaux,
Phys. Rev. Lett. {\bf 78}, 5 (1997).
\bibitem{6}
G.~Bressi, G.~Carugno, R.~Onofrio, and G.~Ruoso,
Phys. Rev. Lett. {\bf 88}, 041804 (2002).
\bibitem {7}
U.~Mohideen and A.~Roy,
{ Phys. Rev. Lett.} {\bf 81}, 4549 (1998).
\bibitem {8}
G.~L.~Klimchitskaya, A.~Roy, U.~Mohideen, and V.~M.~Mostepanenko,
{Phys. Rev. A} {\bf 60}, 3487 (1999).
\bibitem {9}
A.~Roy, C.-Y.~Lin, and U.~Mohideen,
{Phys. Rev. D} {\bf 60}, 111101(R) (1999).
\bibitem{10}
B.~W.~Harris, F.~Chen, and U.~Mohideen,
{Phys. Rev. A} {\bf 62}, 052109 (2000).
\bibitem {11}
A.~Roy and U.~Mohideen,
{Phys. Rev. Lett.} {\bf 82}, 4380 (1999).
\bibitem {12}
V.~B.~Bezerra, G.~L.~Klimchitskaya, and V.~M.~Mostepanenko,
{Phys. Rev. A} {\bf 62}, 014102 (2000).
\bibitem{13}
A.~Lambrecht and S.~Reynaud,
Eur. Phys. J. D {\bf 8}, 309 (2000).
\bibitem {14}
G.~L.~Klimchitskaya, U.~Mohideen, and V.~M.~Mostepanenko,
{ Phys. Rev. A}
{\bf 61}, 062107 (2000).
\bibitem{15}
M.~Bordag, G.~L.~Klimchitskaya,
and V.~M.~Mostepanenko,
 { Int. J. Mod. Phys. A} 
{\bf 10}, 2661 (1995).
\bibitem{16}
G.~L.~Klimchitskaya
and Yu.~V.~Pavlov,
 { Int. J. Mod. Phys. A} 
{\bf 11}, 3723 (1996).
\bibitem {17}
V.~B.~Bezerra, G.~L.~Klimchitskaya, and C.~Romero,
 { Phys. Rev.} A
{\bf 61}, 022115 (2000).
\bibitem {18}
C.~Genet, A.~Lambrecht, and S.~Reynaud,
Phys. Rev. A {\bf 62}, 012110 (2000).
\bibitem{19}
M.~Bordag, B.~Geyer, G.~L.~Klimchitskaya,
and V.~M.~Mostepanenko,
{ Phys. Rev. Lett.}  {\bf 85}, 503  (2000).
\bibitem{20}
G.~L.~Klimchitskaya and V.~M.~Mostepanenko,
{ Phys. Rev. A}  {\bf 63}, 062108  (2001).
\bibitem{21}
B.~Geyer, G.~L.~Klimchitskaya,
and V.~M.~Mostepanenko,
 { Int. J. Mod. Phys. A} 
{\bf 16}, 3291 (2001).
\bibitem {22}
V.~B.~Bezerra, G.~L.~Klimchitskaya, and C.~Romero,
 { Phys. Rev.} A
{\bf 65}, 012111 (2002).
\bibitem {23}
V.~B.~Bezerra, G.~L.~Klimchitskaya,  and V.~M.~Mostepanenko,
 { Phys. Rev.} A
{\bf 65}, 052113 (2002).
\bibitem{24}
B.~Geyer, G.~L.~Klimchitskaya,
and V.~M.~Mostepanenko,
 { Phys. Rev.} A
{\bf 65}, 062109 (2002).
\bibitem{25}
M.~Bordag, U.~Mohideen, and V.~M.~Mostepanenko,
{ Phys. Rep.} {\bf 353}, 1 (2001).
\bibitem{26}
M.~Bordag, B.~Geyer, G.~L.~Klimchitskaya, and V.~M.~Mostepanenko,
{Phys. Rev. D}  {\bf 58}, 075003 (1998).
\bibitem{27}
M.~Bordag, B.~Geyer, G.~L.~Klimchitskaya, and V.~M.~Mostepanenko,
{Phys. Rev. D}  {\bf 60}, 055004 (1999).
\bibitem{28}
J.~C.~Long, H.~W.~Chan, and J.~C.~Price,
Nucl. Phys. B {\bf 539}, 23 (1999).
\bibitem{29}
M.~Bordag, B.~Geyer, G.~L.~Klimchitskaya, and V.~M.~Mostepanenko,
{Phys. Rev. D}  {\bf 62}, 011701(R) (2000).
\bibitem{30}
V.~M.~Mostepanenko and M.~Novello,
{Phys. Rev. D}
{\bf 63}, 115003 (2001).
\bibitem{31}
E.~Fischbach, D.~E.~Krause, V.~M.~Mostepanenko, and M.~Novello,
{Phys. Rev. D}
{\bf 64}, 075010 (2001).
\bibitem{32}
H.~B.~Chan, V.~A.~Aksyuk, R.~N.~Kleiman, D.~J.~Bishop, and F.~Capasso,
Science {\bf 291}, 1941 (2001).
\bibitem{33}
H.~B.~Chan, V.~A.~Aksyuk, R.~N.~Kleiman, D.~J.~Bishop, and F.~Capasso,
Phys. Rev. Lett. {\bf 87}, 211801 (2001).
\bibitem{34}
E.~Buks and M.~L.~Roukes,
{Phys. Rev. B}  {\bf 63}, 033402 (2001).
\bibitem{35}
Yu.~S.~Barash,
Izv. Vuzov. Ser. Radiofiz. {\bf 16}, 1086 (1973)
[Sov. Radiophys. {\bf 16}, 945 (1973)].
\bibitem{36}
S.~J.~van~Enk,
{Phys. Rev. A} {\bf 52}, 2569 (1995).
\bibitem{37}
R.~Golestanian and M.~Kardar,
{Phys. Rev. Lett.} {\bf 78}, 3421 (1997).
\bibitem{38}
R.~Golestanian and M.~Kardar,
{Phys. Rev. A} {\bf 58}, 1713 (1998).
\bibitem{39}
T.~Emig, A.~Hanke, R.~Golestanian, and M.~Kardar,
{Phys. Rev. Lett.} {\bf 87}, 260402 (2001).
\bibitem{40}
L.~S.~Levitov,
Europhys. Lett. {\bf 8}, 499 (1989).
\bibitem{41}
J.~S.~Hoye and I.~Brevik,
Physica A {\bf 181}, 413 (1992).
\bibitem{42}
F.~Chen, U.~Mohideen, G.~L.~Klimchitskaya, and 
V.\ M.\ Mos\-te\-pa\-nen\-ko,
Phys. Rev. Lett. {\bf 88}, 101801 (2002). 
\bibitem{43}
I.~E.~Dzyaloshinskii, E.~M.~Lifshitz, and L.~P.~Pitaevskii,
Usp. Fiz. Nauk {\bf 73}, 381 (1961)
[Sov. Phys. Usp. {\bf 4}, 153 (1961)].
\bibitem {44}
C.~M.~Hargreaves,
Proc. K. Ned. Akad. Wet.
{\bf 68}, 231 (1965).
\bibitem {45}
J.~Schwinger, L.~L.~DeRaad,~Jr., and K.~A.~Milton,
{ Ann. Phys.} (N.Y.) {\bf 115}, 1 (1978).
\bibitem{46}
V.~M.~Mostepanenko and N.~N.~Trunov,
Yad. Fiz. {\bf 42}, 1297 (1985)
[Sov. J. Nucl. Phys. {\bf 42}, 818 (1985)].
\bibitem {47}
V.~B.~Bezerra, G.~L.~Klimchitskaya, and C.~Romero,
 { Int. J. Mod. Phys. A} 
{\bf 16}, 3103 (2001).
\bibitem{48}
J.~Blocki, J.~Randrup, W.~J.~Swiatecki, and C.~F.~Tsang,
Ann. Phys. (N.Y.) {\bf 105}, 427 (1977).
\bibitem{49}
P.~Johansson and P.~Apell, Phys. Rev. B {\bf 56},
4159 (1997).
\bibitem{50}
M.~Schaden and L.~Spruch, 
Phys. Rev. Lett. {\bf 84}, 459 (2000).
\bibitem {50a}
F.~Chen and U.~Mohideen,
Rev. Sci. Instr. {\bf 72}, 3100 (2001).
\bibitem{51}
E.~Fischbach and C.~L.~Talmadge,
{\it The Search for Non-Newtonian Gravity}
(Springer, New York, 1999).
\bibitem{52}
N.~Arkani-Hamed, S.~Dimopoulos, and G.~Dvali,
Phys. Rev. D {\bf 59}, 086004 (1999).
\end{thebibliography}
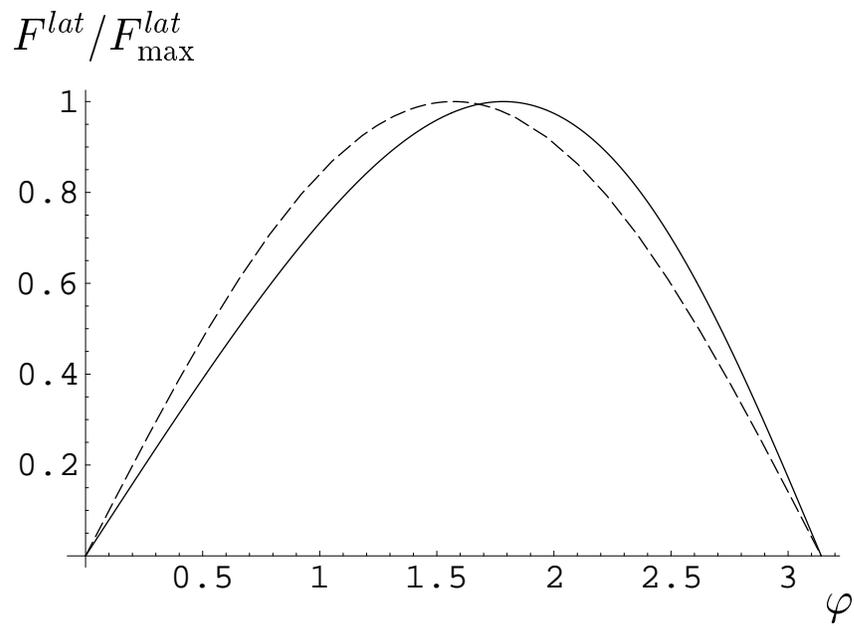
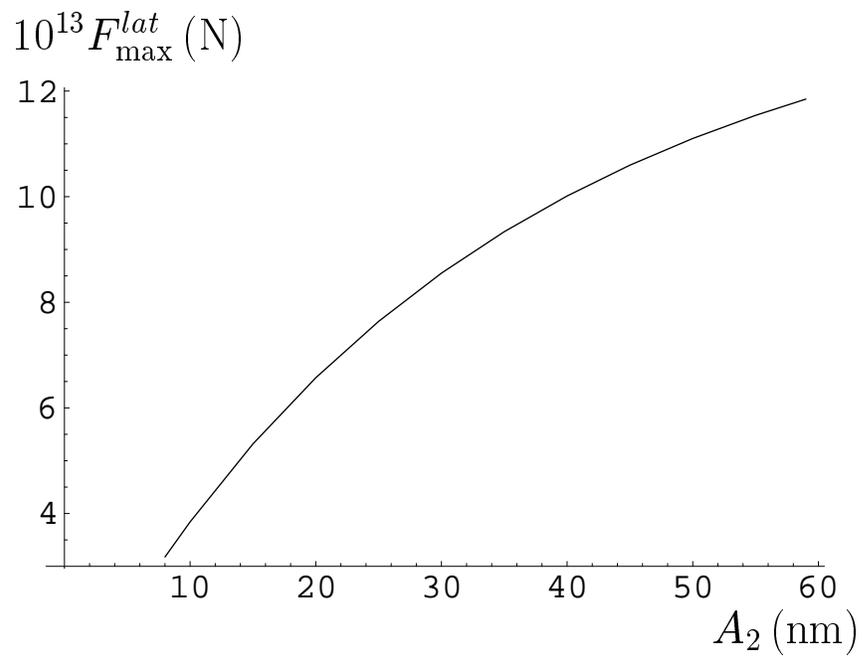
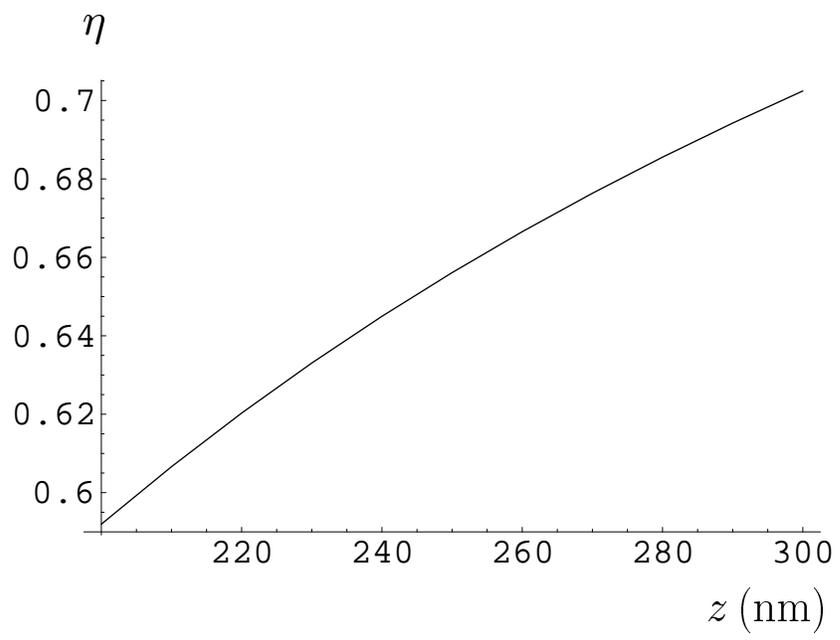
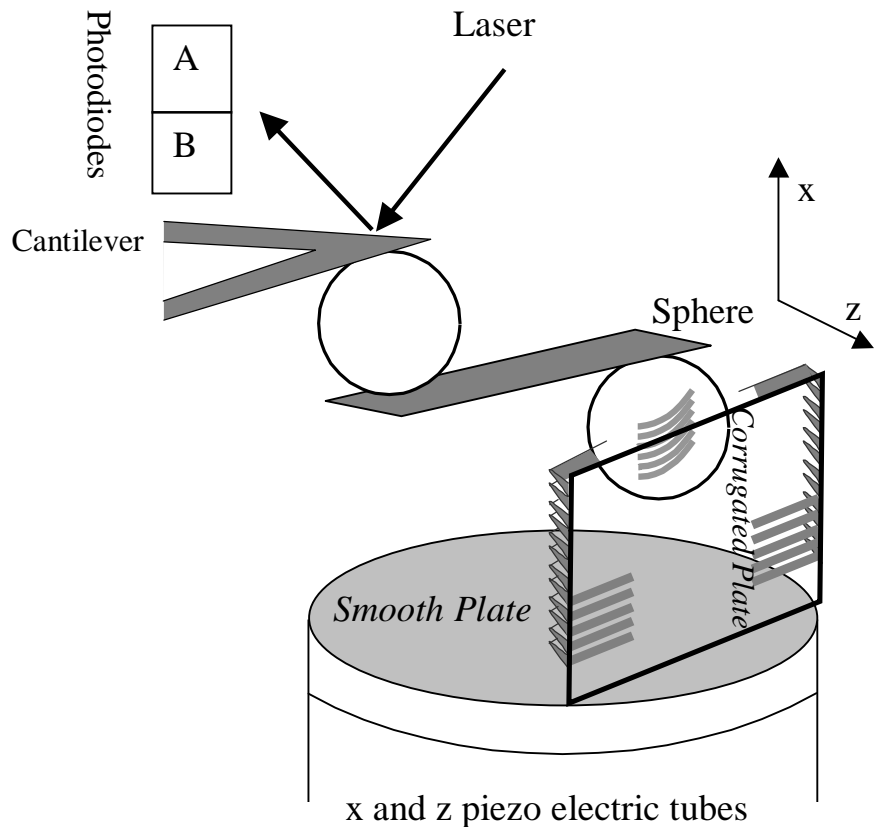
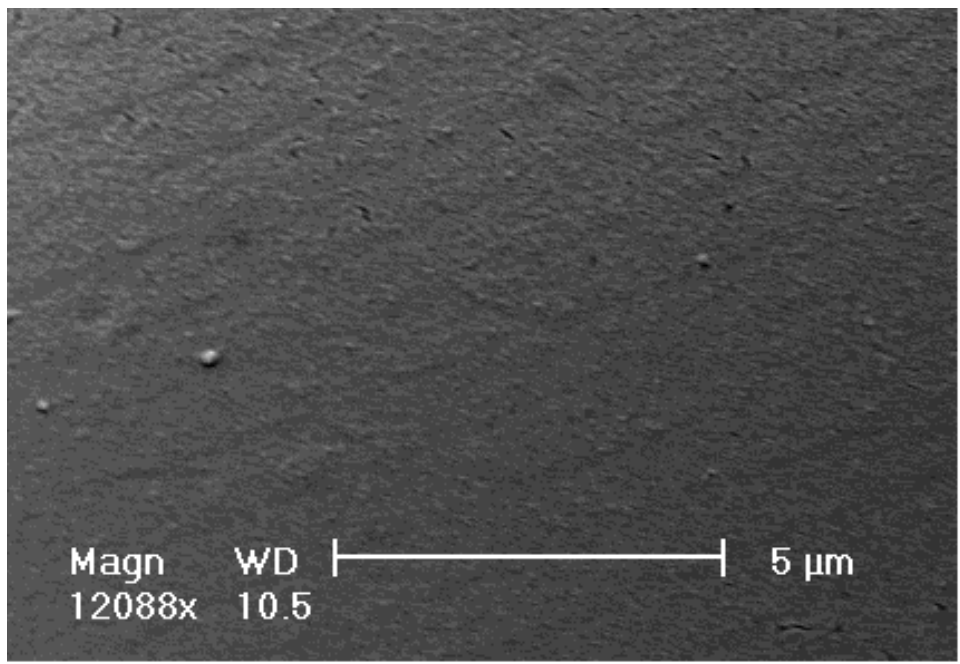
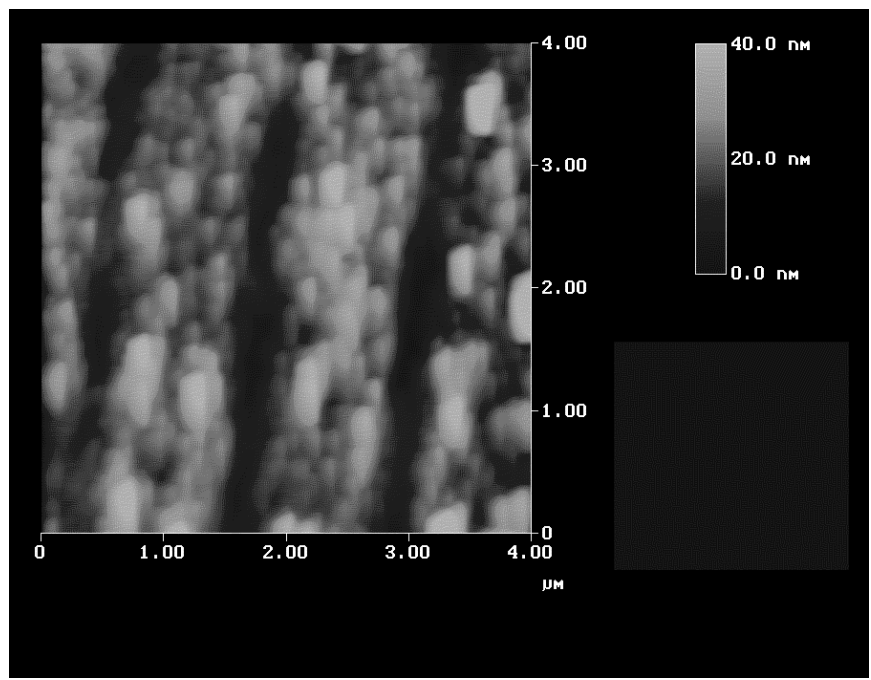
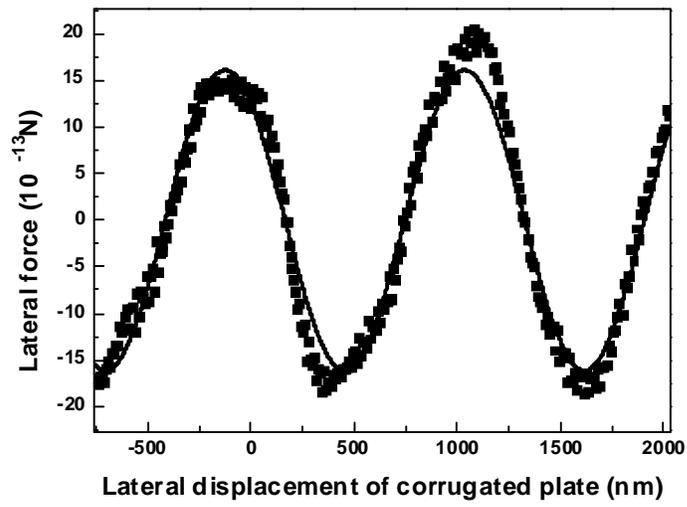
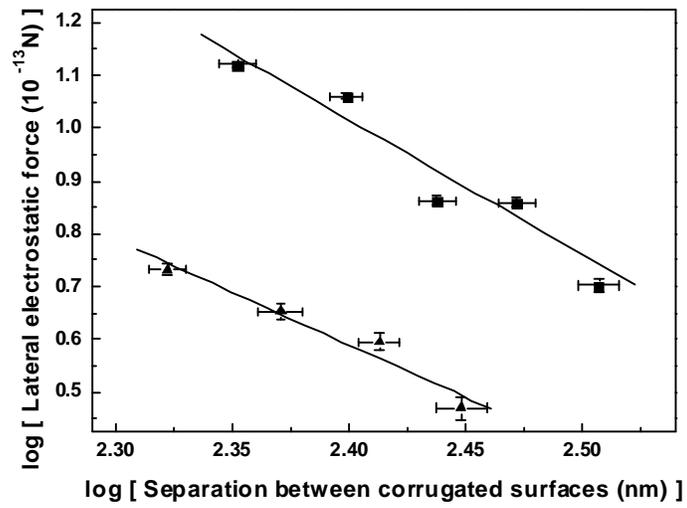
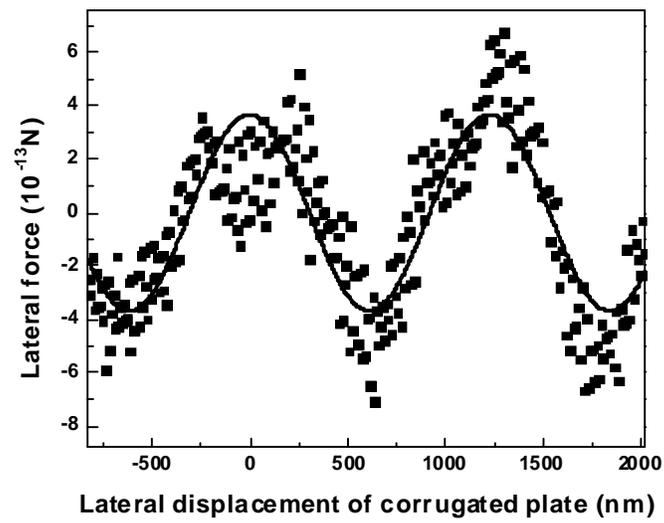
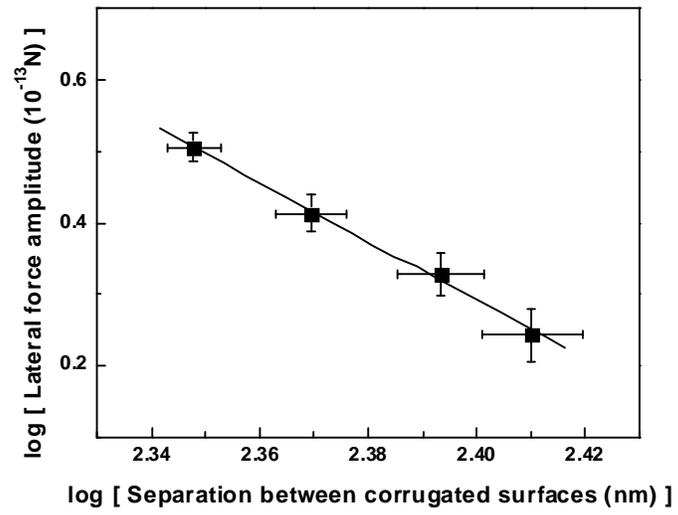
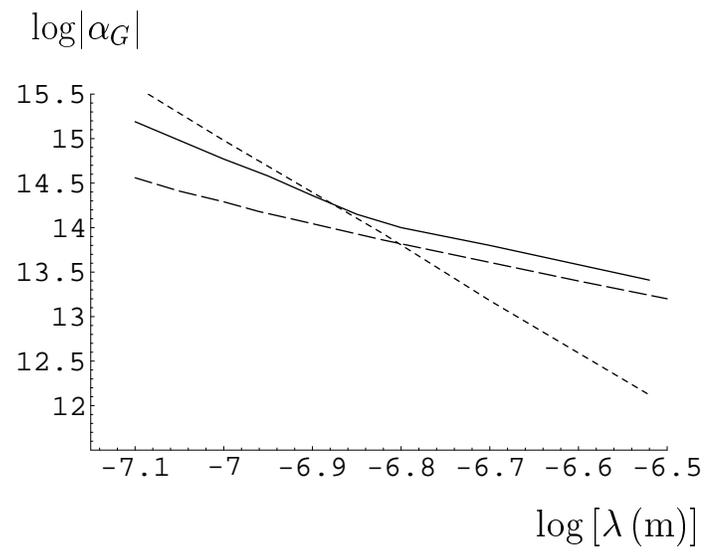
\end{document}